# Optically addressable molecular spins for quantum information processing

S. L. Bayliss[1]†, D. W. Laorenza[2]†, P. J. Mintun[1], B. Diler[1], D. E. Freedman[2]*, D. D. Awschalom[1,3,4]*

[1]Pritzker School of Molecular Engineering, University of Chicago, Chicago, IL 60637, USA.

[2]Department of Chemistry, Northwestern University, Evanston, IL 60208, USA.

[3]Department of Physics, University of Chicago, Chicago, IL 60637, USA.

[4]Center for Molecular Engineering and Materials Science Division, Argonne National Laboratory, Lemont, IL 60439, USA.

†These authors contributed equally to this work.

*Correspondence to: awsch@uchicago.edu, danna.freedman@northwestern.edu

**Spin-bearing molecules are promising building blocks for quantum technologies as they can be chemically tuned, assembled into scalable arrays, and readily incorporated into diverse device architectures. In molecular systems, optically addressing ground-state spins would enable a wide range of applications in quantum information science, as has been demonstrated for solid-state defects. However, this important functionality has remained elusive for molecules. Here, we demonstrate such optical addressability in a series of synthesized organometallic, chromium(IV) molecules. These compounds display a ground-state spin that can be initialized and read out using light, and coherently manipulated with microwaves. In addition, through atomistic modification of the molecular structure, we tune the spin and optical properties of these compounds, paving the way for designer quantum systems synthesized from the bottom-up.**

Optically addressable solid-state spins (*1–4*), are an important platform for quantum information science, with impressive demonstrations ranging from quantum teleportation (*5*) to the mapping of individual nuclear spins (*6*). The optical-spin interface of these solid-state systems is crucial for a diverse range of applications, from nanoscale sensing to long-distance quantum communication, as it enables straightforward single-spin readout and initialization. However, for this family of qubits, synthetic tunability of optical and spin properties, deterministic fabrication of multi-qubit arrays, or translation of spin centers between different host materials and devices remain outstanding goals.

By contrast, chemical synthesis of molecular spin systems affords bottom-up qubit design (*7*). A chemical approach offers: tunability through atomistic control over the qubit; scalability via chemical assembly of extended structures; and portability across different environments (e.g. solution, surface, solid-state), since the qubit is not confined to a specific host. These capabilities provide remarkable control over the intrinsic and extrinsic environment of molecular qubits. Notably, with chemical synthesis, nuclear spins can be controllably placed around a molecular qubit (*8*), arrays of spins can be created in 1-, 2- and 3-dimensional architectures (*9*, *10*), and molecular spins can be integrated into electronic and photonic devices (*11*, *12*). Molecular systems have shown impressive demonstrations including long spin coherence (*13*), manipulation of photoexcited triplet states (*14–16*), and quantum optics with spin-singlet ($S$=0) organic molecules (*17*). However, in contrast to spins in semiconductors, the ground-state spin of molecular systems has lacked an optical-spin interface for both qubit initialization and readout (*18*). Creating such an interface in a molecular platform would generate a class of qubits which can be engineered with atomic precision, with transformative applications for bottom-up quantum technologies ranging from quantum sensors to hybrid quantum systems.

Here, through bottom-up design, we synthesize a series of tunable molecular qubits with such an optically addressable ground-state spin. We show that these molecular spin qubits can be initialized and read out with light, and coherently manipulated with microwave fields. By tuning both their optical and spin properties through control of molecular structure, we demonstrate the power of bottom-up qubit creation.

To achieve the desired optical addressability, we target a tunable molecular system consisting of a metal ion bonded to organic moieties (ligands), comprising a portable qubit of ~1 nm size. This organometallic motif provides a well-defined qubit through the electronic spin of the central metal ion, with highly controllable ligands surrounding the metal to offer synthetic tunability.

The key requirements for such an optically addressable molecular spin qubit are: (*i*) a ground-state spin which can be coherently manipulated, and (*ii*) a spin-selective optical process to initialize and read out the spin. To achieve these functionalities, we selected a chromium ion ($Cr^{4+}$) coordinated by strong-field (aryl) ligands in a high-symmetry configuration, which gives rise to the energy-level structure shown in Fig. 1A. The $d^2$ electronic configuration of $Cr^{4+}$ in a pseudo-tetrahedral environment produces a spin-triplet ($S$=1) ground state with a small ground-state zero-field splitting, characterized by the parameters $D$ and $E$, allowing for spin manipulation at readily available microwave frequencies.

A strong ligand-field environment ensures that the lowest lying electronic excited state is a spin-singlet ($S$=0) (*19*). This configuration leads to narrow optical transitions between the $S$=1 ground state and the $S$=0 excited state, which when combined with the ground-state zero-field splitting, enables optical spin readout and initialization (i.e. spin polarization) through spin-selective resonant excitation. First, optical readout of the ground-state spin is possible since a probed spin sublevel (e.g. $|0\rangle$ in Fig. 1A), will give rise to more photoluminescence (PL) than the other spin sublevels (e.g. $|\pm 1\rangle$ in Fig. 1A). Second, optical polarization of the ground-state spin results when selective excitation, combined with spontaneous emission, transfers population from the probed to the other spin sublevels (*20*). This is referred to as optical pumping or hole burning. Importantly, to accumulate spin polarization over multiple excitation and emission cycles, the ground-state spin-lattice relaxation time ($T_1$) must be much longer than the excited-state lifetime ($T_{opt}$). These components are the key ingredients we use to obtain the desired optical-spin interface.

With these criteria in mind, we synthesized the three $Cr^{4+}$ compounds (Fig. 1B), which differ by the placement of a single $CH_3$ (methyl group) on the coordinating ligands, through solution-phase chemistry. In brief, we react the appropriate aryl lithium species with $Cr^{3+}Cl_3(THF)_3$ at −78°C which undergoes a disproportionation or auto-oxidation to the corresponding tetrahedral $Cr^{4+}R_4$ (R=*o*-tolyl, 2,3-dimethylphenyl, 2,4-dimethylphenyl, see supplementary materials for further details) (*21*). We diluted each compound in their $S = 0$ isostructural tin analogues to form dilute molecular crystals (**1, 2, 3**), illustrated in Fig. 1C, thus reducing interactions between $Cr^{4+}$ centers. All experiments were performed on **1**–**3** in an optical cryostat with microwave access (≈4-5 K at the sample mount, Fig. 1C and supplementary materials) unless otherwise stated.

Under off-resonant excitation (785 nm), ground-state population is promoted to the first $S$=1 excited state, undergoes fast intersystem crossing to the $S$=0 state, and decays to the $S$=1 ground state, emitting near-infrared PL. For **1**–**3**, this emission comprises sharp zero-phonon lines (ZPLs) ranging from 1009 – 1025 nm (Fig. 1D), along with longer-wavelength phonon sidebands. The minor ligand modifications in **1–3** also result in unique ground-state spin structure, as observed in ground state electron spin resonance (ESR) measurements (Fig. 1G), with $D$ and $E$ lying in the readily addressable regime of <5 GHz for each compound (we take $D, E > 0$ - see supplementary materials). These features, along with optical lifetimes (3.3–6.9 µs, Fig. 1F) that are much shorter than $T_1$ (see below and supplementary materials), therefore suggest that **1–3** satisfy the above criteria for optically addressable molecular qubits, with synthetically tunable optical and spin properties.

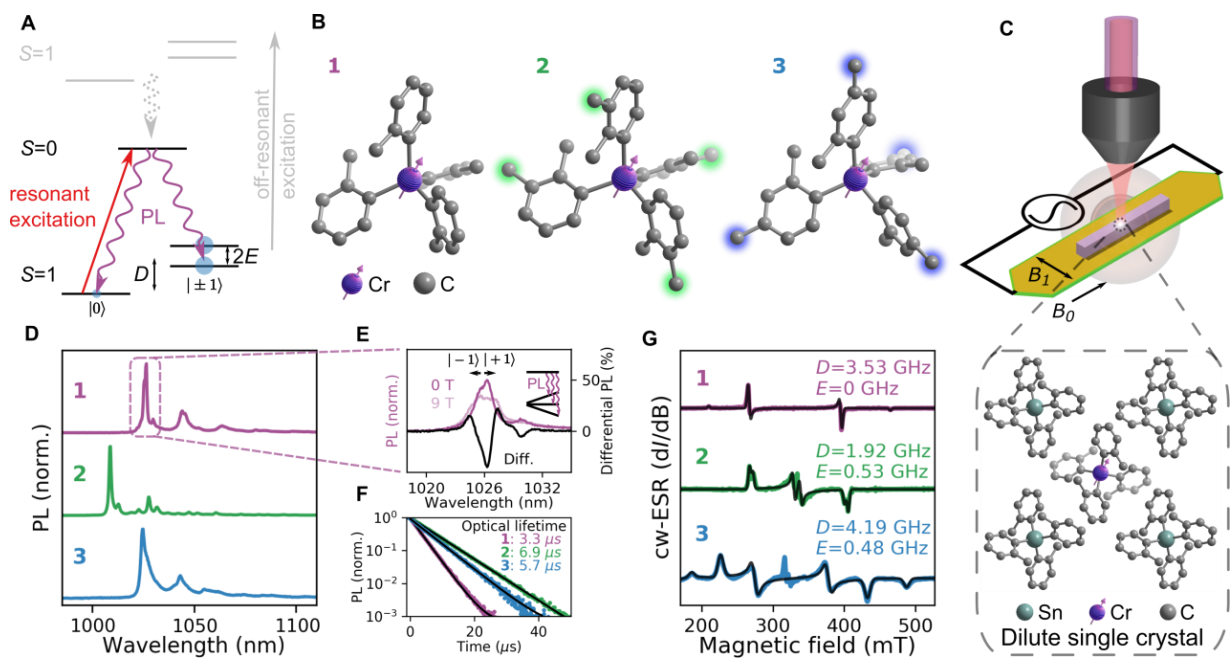

**Fig. 1. Generating an optical interface for ground-state molecular spin qubits.** (**A**) Energy-level diagram of $Cr^{4+}$ in **1–3** depicting photoluminescence (PL) from the $S=0$ state. (**B**) Molecular structures for **1–3** determined by single-crystal X-ray diffraction. Hydrogen atoms are omitted for clarity. Ligand modifications for **2** and **3** are highlighted in green and blue. Chromium and carbon atoms are shown in purple and gray, respectively. (**C**) Experimental schematic depicting optical excitation and PL collection for spin initialization and readout. Each $Cr^{4+}$ compound is diluted in a single crystal (purple) of the isostructural $S = 0$ tin (Sn) analogue. An illustrative structure is shown. A microwave field ($B_1$) from a waveguide (gold) is used for spin manipulation, and a static field ($B_0$) enables Zeeman splitting. (**D**) PL spectra for **1–3** at 4 K using off-resonant (785 nm) excitation. (**E**) Zeeman splitting of the zero-phonon line of **1** at 9 Tesla. (**F**) Optical lifetimes for **1–3** measured using resonant excitation at the zero-phonon line. (**G**) X-band continuous-wave electron spin resonance (cwESR) spectra for **1–3** collected at 77 K. Simulations are shown in black, along with extracted $D$ and $E$ parameters. The central resonances at $g \approx 2$-2.1 are discussed in the supplementary materials.

To further confirm the level structure in Fig. 1A, we measure the emission of **1** under a high magnetic field using off-resonant excitation (Fig. 1E). Due to the $S=0$ excited state, the Zeeman splitting of the ground state manifests directly as a shift in the optical emission energies. This effect is clearly shown by taking the difference in PL spectra at 9 and 0 T: optical emission into the $|\pm 1\rangle$ spin sublevels shift to lower and higher energies, giving characteristic peaks on either side
3

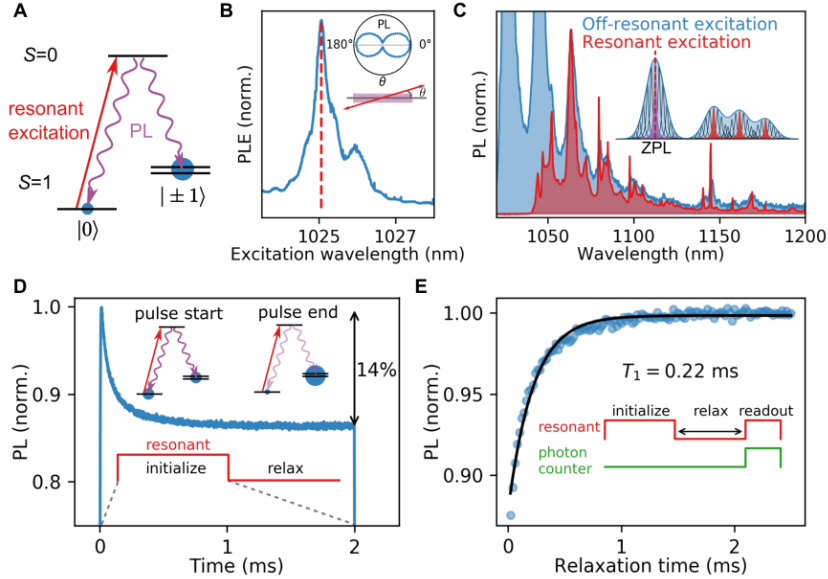

**Fig. 2. All-optical ground-state spin initialization and readout of 1.** (**A**) Energy-level structure showing optical spin initialization through spin-selective excitation. (**B**) Photoluminescence excitation (PLE) spectrum obtained by sweeping a narrow-line laser over the zero-phonon line. The dashed line shows the excitation wavelength used for all following experiments. Inset: dependence of the PL on laser polarization, defined by the angle θ from the crystal long axis. (**C**) Phonon sideband under resonant and off-resonant excitation showing emission line narrowing. Inset: schematic of subensemble excitation. (**D**) Time-resolved optical spin initialization. (**E**) All-optical measurement of the spin-lattice relaxation time ($T_1$).

of the zero-field ZPL in the differential spectrum, along with a central dip (the feature at 1030 nm arises from the vibrational sideband, see supplementary materials).

To demonstrate an optical-spin interface in these systems, we now focus on **1** as an illustrative example before discussing **2** and **3**. Using a narrow-line laser, we resonantly excite the $S$=1 ground state to the $S$=0 excited state (Fig. 2A) and collect emission into the phonon sideband to remove excitation laser scatter. First, we characterize the emission as a function of the excitation wavelength (Fig. 2B), showing a ZPL at 1025 nm: we excite at this ZPL maximum (dashed line Fig 2B) for all following experiments. To further maximize emission, we align the excitation polarization with the optical dipole transition, which is collinear with the long axis of the crystal (Fig. 2B inset). While the optical inhomogeneous linewidth of ≈150 GHz shown in Fig. 2B appears prohibitive for spin-selective excitation, as this linewidth is >> $D$, resonant excitation addresses a narrower subensemble of molecules from the inhomogeneous distribution (likely broadened by strain) (*22*). To demonstrate that this subensemble linewidth is indeed much narrower than the inhomogeneous linewidth, we compare the phonon sidebands under resonant excitation and off-resonant excitation (Fig. 2C). The emission line narrowing (*23*) under resonant excitation indicates that the ensemble ZPL indeed consists of narrower subensembles, which we use for all following spin-selective experiments.

We next measure all-optical initialization and readout of the ground-state spin using hole-burning and recovery. To initialize the spin, we apply the pulse sequence outlined in Fig. 2D consisting of a long optical pulse (2 ms), followed by a wait time to equilibrate ground-state spin populations before the next pulse. The emission during the optical pulse shows the characteristic behavior of optical spin polarization: a gradual drop in emission as population is pumped from the probed ground-state spin sublevel (the 'bright' state) and into the other ('dark') spin sublevels. The optical contrast between the start and the end of the pulse places a lower bound on the spin polarization of 14% (see supplementary materials).

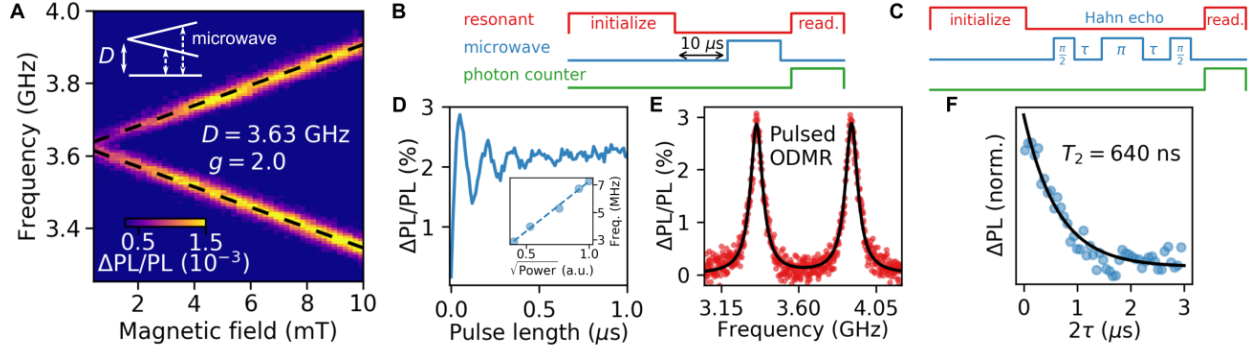

**Fig. 3. Optically detected magnetic resonance (ODMR) and coherent spin manipulation of the ground state of 1.** (**A**) ODMR as a function of magnetic field and microwave frequency using continuous-wave optical excitation. Dashed lines are a simulation with the stated $g$ and $D$ values. (**B**) Pulsed ODMR and (**C**) Hahn-echo sequences. (**D**) Rabi oscillations between the $|0\rangle$ and $|-1\rangle$ spin sublevels ($B_0$=10 mT). Inset: microwave-power dependence of the Rabi oscillation frequency. (**E**) Pulsed ODMR spectrum ($B_0$=10 mT) and double Lorentzian fit (black). (**F**) Optically detected ground-state spin coherence ($B_0$=2 mT) with exponential fit (black).

Using this spin initialization, we now measure the ground-state spin-lattice relaxation time, $T_1$ by performing the two-pulse experiment outlined in Fig. 2E. This sequence consists of an initialization pulse (300 µs), a variable relaxation time and a readout pulse (20 µs). The initialization pulse transfers population to the 'dark' spin sublevels. As ground-state spin population relaxes back to the 'bright' sublevel, the emission increases. Measuring this emission at variable relaxation times yields $T_1$=0.22(1) ms. Since $T_1$ is significantly longer than the optical lifetime ($T_{opt}$=3.3 µs, Fig. 1F), this confirms that many optical cycles can be used to accumulate ground-state spin polarization.

We next manipulate the ground-state spin of **1** using a microwave field. First, using continuous wave (cw) optical excitation, we place a subensemble of spins in the 'dark' state and monitor changes in emission (ΔPL) as we sweep the microwave frequency. When this microwave frequency matches the spin sublevel splitting, the 'dark' and 'bright' sublevels are mixed, resulting in increased PL. Fig. 3A shows this optically detected magnetic resonance (ODMR) as a function of both the microwave frequency and an external magnetic field applied along the long axis of the crystal. The zero-field cw-ODMR spectrum provides $D$=3.63 GHz, while the Zeeman splitting yields a $g$-factor of 2.0, in agreement with the ESR measurements (Fig. 1G).

To demonstrate coherent control over the ground-state spin, we drive Rabi oscillations (Fig. 3D) using the pulsed ODMR sequence outlined in Fig. 3B. This sequence consists of an optical initialization pulse, a wait time, a variable length microwave pulse, and an optical readout pulse. The inset shows the expected square-root dependence of the Rabi frequency on the applied microwave power. Next, using a π-pulse calibrated from Fig. 3D, we perform pulsed ODMR at a fixed magnetic field, $B_0$=10 mT, while varying the microwave frequency (Fig. 3E). Finally, by replacing the single microwave pulse in Fig. 3B with a Hahn-echo sequence (Fig. 3C), we measure the spin coherence time $T_2$=640(60) ns (Fig. 3F, $B_0$=2 mT). The final π/2-pulse in the sequence projects the coherences onto spin populations for optical readout. Importantly, in these pulsed ODMR experiments, the wait time (10 µs~$3T_{opt}$) between initialization and microwave manipulation ensures population is in the ground state prior to coherent control. This wait time, along with the above agreement between the ODMR and ESR spin parameters, verifies that we coherently control the ground-state spin. Furthermore, the measured $T_2$, likely limited by the surrounding hydrogen nuclear spins, is comparable to other organometallic systems in nuclear spin-rich environments (*13*, *24*). Thus, with **1**, we demonstrate optical initialization, microwave coherent control and optical readout of the ground-state spin in a molecular qubit.

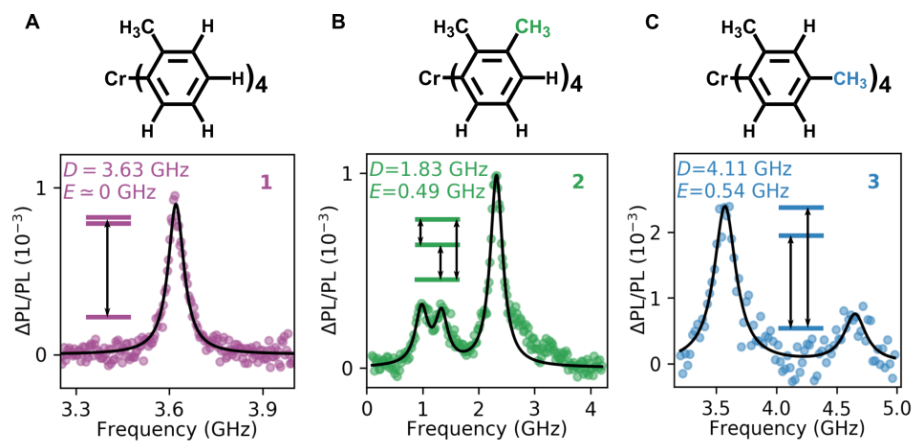

**Fig. 4. Optical spin addressability with synthetic tunability.** (**A, B, C**) cw-ODMR spectra and simulations (black) for **1-3,** with microwave transitions and ligand modifications depicted.

Having demonstrated an optical-spin interface and coherent spin control for **1**, we now highlight how this functionality can be translated to a much broader class of molecules. In Fig. 4, we show robust optical initialization, microwave spin manipulation and optical readout of **2** and **3** through cw-ODMR. As captured by the simulations, the variable peak intensities arise from ESR selection rules (see supplementary materials). These results therefore demonstrate engineered optical-spin interfaces in a bottom-up system with synthetic control over magnetic, electronic, and physical structure. Chemical design provides an immediate pathway to enhance such systems. For example, chemical replacement of the hydrogen nuclei (Fig. 4A) around the metal center in **1** through deuteration should significantly enhance the electronic spin coherence (*24*, *25*). Furthermore, the generation of a significant $E$ in compounds **2** and **3** exemplifies the ability to engineer noise-insensitive (i.e. clock-like) transitions by reducing symmetry in molecular architectures (*26*). This work demonstrates that bottom-up design may be harnessed to create a range of quantum systems, such as scalable arrays of qubits patterned on surfaces, with variable optical and microwave resonances for single-spin addressability. Alternatively, by deterministically placing nuclear spins around the metal center, tailor-made, long-lived registers with an optical interface could be created (*27*). Finally, the portability and nanometer-scale of molecular qubits holds promise for their integration with diverse systems ranging from optical cavities for quantum optical networking to biological macromolecules for nanoscale sensing (*17*, *28*, *29*). These results open pathways to design and create quantum technologies from the bottom-up.

**References**

1. M. Atatüre, J. Dreiser, A. Badolato, A. Högele, K. Karrai, A. Imamoglu, Quantum-dot spin-state preparation with near-unity fidelity. *Science*. **312**, 551–553 (2006).
2. D. D. Awschalom, R. Hanson, J. Wrachtrup, B. B. Zhou, Quantum technologies with optically interfaced solid-state spins. *Nat. Photonics*. **12**, 516–527 (2018).
3. A. Gottscholl, M. Kianinia, V. Soltamov, S. Orlinskii, G. Mamin, C. Bradac, C. Kasper, K. Krambrock, A. Sperlich, M. Toth, I. Aharonovich, V. Dyakonov, Initialization and read-out of intrinsic spin defects in a van der Waals crystal at room temperature. *Nat. Mater.* (2020).
4. N. Chejanovsky, A. Mukherjee, Y. Kim, A. Denisenko, A. Finkler, T. Taniguchi, K. Watanabe, D. B. R. Dasari, J. H. Smet, J. Wrachtrup, Single spin resonance in a van der Waals embedded paramagnetic defect. *arXiv:1906.05903* (2019).
5. W. Pfaff, B. J. Hensen, H. Bernien, S. B. Van Dam, M. S. Blok, T. H. Taminiau, M. J. Tiggelman, R. N. Schouten, M. Markham, D. J. Twitchen, R. Hanson, Unconditional quantum teleportation between distant solid-state quantum bits. *Science*. **345**, 532–535 (2014).
6. M. H. Abobeih, J. Randall, C. E. Bradley, H. P. Bartling, M. A. Bakker, M. J. Degen, M. Markham, D. J. Twitchen, T. H. Taminiau, Atomic-scale imaging of a 27-nuclear-spin cluster using a quantum sensor. *Nature*. **576**, 411–415 (2019).
7. A. Gaita-Ariño, F. Luis, S. Hill, E. Coronado, Molecular spins for quantum computation. *Nat. Chem.* **11**, 301–309 (2019).
8. C. E. Jackson, C. Y. Lin, S. H. Johnson, J. Van Tol, J. M. Zadrozny, Nuclear-spin-pattern control of electron-spin dynamics in a series of V(IV) complexes. *Chem. Sci.* **10**, 8447–8454 (2019).
9. T. Yamabayashi, M. Atzori, L. Tesi, G. Cosquer, F. Santanni, M. E. Boulon, E. Morra, S. Benci, R. Torre, M. Chiesa, L. Sorace, R. Sessoli, M. Yamashita, Scaling Up Electronic Spin Qubits into a Three-Dimensional Metal-Organic Framework. *J. Am. Chem. Soc.* **140**, 12090–12101 (2018).
10. A. Urtizberea, E. Natividad, P. J. Alonso, M. A. Andrés, I. Gascón, M. Goldmann, O. Roubeau, A Porphyrin Spin Qubit and Its 2D Framework Nanosheets. *Adv. Funct. Mater.* **28**, 1801695 (2018).
11. R. Vincent, S. Klyatskaya, M. Ruben, W. Wernsdorfer, F. Balestro, Electronic read-out of a single nuclear spin using a molecular spin transistor. *Nature*. **488**, 357–360 (2012).
12. M. Oxborrow, J. D. Breeze, N. M. Alford, Room-temperature solid-state maser. *Nature*. **488**, 353–356 (2012).
13. J. M. Zadrozny, J. Niklas, O. G. Poluektov, D. E. Freedman, Millisecond coherence time in a tunable molecular electronic spin qubit. *ACS Cent. Sci.* **1**, 488–492 (2015).
14. J. Wrachtrup, C. Von Borczyskowski, J. Bernard, M. Orritt, R. Brown, Optical detection of magnetic resonance in a single molecule. *Nature*. **363**, 244–245 (1993).
15. J. Köhler, J. A. J. M. Disselhorst, M. C. J. M. Donckers, E. J. J. Groenen, J. Schmidt, W. E. Moerner, Magnetic resonance of a single molecular spin. *Nature*. **363**, 242–244 (1993).
16. V. Filidou, S. Simmons, S. D. Karlen, F. Giustino, H. L. Anderson, J. J. L. Morton, Ultrafast entangling gates between nuclear spins using photoexcited triplet states. *Nat. Phys.* **8**, 596–600 (2012).
17. D. Wang, H. Kelkar, D. Martin-Cano, D. Rattenbacher, A. Shkarin, T. Utikal, S. Götzinger, V. Sandoghdar, Turning a molecule into a coherent two-level quantum system.

18. M. Atzori, R. Sessoli, The Second Quantum Revolution: Role and Challenges of Molecular Chemistry. *J. Am. Chem. Soc.* **141**, 11339–11352 (2019).
19. B. N. Figgis, M. A. Hitchman, in *Ligand Field Theory and Its Applications* (Wiley-VCH, New York, 2000), pp. 131–141.
20. W. F. Koehl, B. Diler, S. J. Whiteley, A. Bourassa, N. T. Son, E. Janzén, D. D. Awschalom, Resonant optical spectroscopy and coherent control of $Cr^{4+}$ spin ensembles in SiC and GaN. *Phys. Rev. B*. **95**, 035207 (2017).
21. S. U. Koschmieder, B. S. McGilligan, G. McDermott, J. Arnold, G. Wilkinson, B. Hussain-Bates, M. B. Hursthouse, Aryl and aryne complexes of chromium, molybdenum, and tungsten. X-Ray crystal structures of [Cr(2-MeC6H4)(μ-2-MeC6H4)(PMe3)]2, Mo(η2-2-MeC6H3)(2-MeC6H4)2(PMe2Ph)2, and W(η2-2,5-Me2C6H2)(2,5-Me2C6H3)2-(PMe3)2. *J. Chem. Soc. Dalt. Trans.*, 3427–3433 (1990).
22. H. Riesen, Hole-burning spectroscopy of coordination compounds. *Coord. Chem. Rev.* **250**, 1737–1754 (2006).
23. H. Riesen, E. Krausz, Persistent spectral hole-burning, luminescence line narrowing and selective excitation spectroscopy of the R lines of Cr(III) tris(2,2′-bipyridine) in amorphous hosts. *J. Chem. Phys.* **97**, 7902–7910 (1992).
24. A. Ardavan, O. Rival, J. J. L. Morton, S. J. Blundell, A. M. Tyryshkin, G. A. Timco, R. E. P. Winpenny, Will spin-relaxation times in molecular magnets permit quantum information processing? *Phys. Rev. Lett.* **98**, 057201 (2007).
25. A. Ardavan, A. M. Bowen, A. Fernandez, A. J. Fielding, D. Kaminski, F. Moro, C. A. Muryn, M. D. Wise, A. Ruggi, E. J. L. McInnes, K. Severin, G. A. Timco, C. R. Timmel, F. Tuna, G. F. S. Whitehead, R. E. P. Winpenny, Engineering coherent interactions in molecular nanomagnet dimers. *npj Quantum Inf.* **1**, 15012 (2015).
26. M. Shiddiq, D. Komijani, Y. Duan, A. Gaita-Ariño, E. Coronado, S. Hill, Enhancing coherence in molecular spin qubits via atomic clock transitions. *Nature*. **531**, 348–351 (2016).
27. M. V. Gurudev Dutt, L. Childress, L. Jiang, E. Togan, J. Maze, F. Jelezko, A. S. Zibrov, P. R. Hemmer, M. D. Lukin, Quantum register based on individual electronic and nuclear spin qubits in diamond. *Science*. **316**, 1312–1316 (2007).
28. F. Shi, Q. Zhang, P. Wang, H. Sun, J. Wang, X. Rong, M. Chen, C. Ju, F. Reinhard, H. Chen, J. Wrachtrup, J. Wang, J. Du, Single-protein spin resonance spectroscopy under ambient conditions. *Science*. **347**, 1135–1138 (2015).
29. R. Chikkaraddy, B. De Nijs, F. Benz, S. J. Barrow, O. A. Scherman, E. Rosta, A. Demetriadou, P. Fox, O. Hess, J. J. Baumberg, Single-molecule strong coupling at room temperature in plasmonic nanocavities. *Nature*. **535**, 127–130 (2016).

**Acknowledgments:** The authors thank M. S. Fataftah for experimental suggestions and insightful discussions, and C. P. Anderson, A. Bourassa, P. Deb, G. Smith, L. R Weiss, M. J. Amdur, K. A. Collins and M. K. Wojnar for helpful comments on the manuscript. We thank P. H. Oyala for technical support with the ESR measurements and R. A. Sponenburg for technical support with ICP-OES experiments. **Funding:** We acknowledge funding from ONR N00014-17-1-3026 and the MRSEC Shared User Facilities at the University of Chicago (NSF DMR-1420709). This work made use of the Caltech EPR facility, which is supported by the NSF (NSF-1531940) and the Dow Next Generation Educator Fund, and IMSERC at Northwestern University, which has received support from Northwestern University, the State of Illinois, and the Int. Institute of Nanotechnology. Metal analysis was performed at the Northwestern University Quantitative Bio-element Imaging Center. D.E.F. and D.W.L. gratefully acknowledge DE-SC0019356 for funding.
**Author contributions:** All authors helped to design the research, perform the research and write the paper. **Competing interests:** A patent application has been filed relating to this work. P.J.M. is a paid consultant to ARCH Venture Partners. **Data and materials availability:** All data are available upon request to the corresponding authors.